\documentclass[12pt,preprint]{aastex} 

\shorttitle{ Hot Emission from Quiescent Regions }
\shortauthors{Schmelz et al.}

\begin{document}

\title{Hinode X-Ray Telescope Detection of Hot Emission from Quiescent Active Regions: A Nanoflare Signature?}

\author{
J.T. Schmelz\altaffilmark{1}, 
S.H. Saar\altaffilmark{2}, 
E.E. DeLuca\altaffilmark{2}, 
L. Golub\altaffilmark{2}, 
V.L. Kashyap\altaffilmark{2}, 
M.A. Weber\altaffilmark{2},
J.A. Klimchuk\altaffilmark{3}
}
\altaffiltext{1}{Physics Department, University of Memphis, Memphis, TN 38152, jschmelz@memphis.edu}
\altaffiltext{2}{Harvard-Smithsonian Center for Astrophysics, 60 Garden St., Cambridge, MA 02138}
\altaffiltext{3}{NASA-GSFC, Code 671.0, Greenbelt, MD 20771}

\begin{abstract}
The X-Ray Telescope (XRT) on the Japanese/USA/UK {\it Hinode (Solar-B)} spacecraft has detected emission from a quiescent active region core that is consistent with nanoflare heating. The fluxes from 10 broadband X-ray filters and filter combinations were used to constructed Differential Emission Measure (DEM) curves. In addition to the expected active region peak at Log T $=$ 6.3-6.5, we find a high-temperature component with significant emission measure at Log T $>$ 7.0. This emission measure is weak compared to the main peak -- the DEM is down by almost three orders of magnitude -- which accounts of the fact that it has not been observed with earlier instruments. It is also consistent with spectra of quiescent active regions: no Fe XIX lines are observed in a CHIANTI synthetic spectrum generated using the XRT DEM distribution. The DEM result is successfully reproduced with a simple two-component nanoflare model. 
\end{abstract}


\section{Introduction}
 
If nanoflares heat the corona (Parker 1983, 1988; Klimchuk 2006), then we would expect an ever-present hot component with a temperature of 10 MK or higher, even in quiescent active regions. A glimpse of this hot material may have been seen by, for example, the Hard X-Ray Imaging Spectrometer on Solar Maximum Mission (Martens, van den Oord \& Hoyng 1985) or the Mg XII imager onboard CORONAS-F (Grechnev et al. 2006). For the most part, however, it has only been possible to put an observational upper limit on the amount of this hot material because high temperature spectral lines like Ca XIX (Lemen, Bentley \& Sylwester 1986; Fludra \& Schmelz 1999) and Fe XIX (Schmelz 1993; Wang, Innes \& Solanki 2006; Teriaca et al. 2006) were not seen except during solar flares. Even observations of the pervasive S XV line by the Yohkoh Bragg Crystal Spectrometer (Culhane et al. 1991) could be explained with cooler (3-4 MK) plasma (Watanabe et al. 1995; Sterling 1997).

Broadband imagers like the Yohkoh Soft X-Ray Telescope (Tsuneta et al. 1991) were designed to investigate hot, flaring plasma, but results tended to rely on filter ratios and an isothermal approximation to estimate plasma parameters. In this letter, we use data from the X-Ray Telescope (XRT; Golub et al. 2007) on Hinode (Kosugi et al. 2007). XRT is a broadband instrument similar to the Soft X-Ray Telescope, but with better spatial resolution, sensitivity, and temperature coverage. A detailed 3 MK thermal map of an active region observed by XRT has already been published by Reale et al. (2007) using an innovative combined filter ratio method. The properties of the instrument allow us to go one step further, however, and perform a true multithermal analysis. XRT should be able to observe the high temperature plasma predicted by nanoflare models, or at least determine a strict upper limit for the amount of emitting material.

\section{Analysis}

XRT observed AR 10955 (S09W30) on 2007 May 13 at 1800 UT. The data set includes full Sun, full resolution images in all filters as well as two filter combinations. The data were processed with standard XRT software available in SolarSoft (xrt\_prep.pro). This includes subtraction of a model dark/read-noise frame, correction for vignetting, removal of high frequency pattern noise, and normalization by exposure time. Spacecraft jitter was removed (xrt\_jitter.pro), and long and short exposures (see Table 1) were co-aligned and combined to increase image dynamic range. The data from the thickest channel (Be\_thick) were subject to additional Fourier filtering to remove some low-level, residual, longer-wavelength noise patterns. We have used updated filter calibrations (Narukage et al. 2009) and 1640 \AA\ of diethylhexyl phthalate, a time-dependent contamination layer on the CCD, which can now be modeled using the make\_xrt\_wave\_resp.pro and make\_xrt\_temp\_resp.pro routines available in SolarSoft.

Figure 1 shows the XRT Ti\_poly image of AR 10955, which produced a small GOES A7 flare at 18:12 UT. Our observations indicate that the effects of this flare were limited to the left side of the active region, so we focused our attention on the outlined area. Even here, well away from the flaring material, there are significant counts in the Be\_thick channel. Since we are specifically searching for hot plasma, we restricted our analysis to those pixels within the outlined area having significant counts in Be\_thick. The noise in the Be\_thick image (on the disk but outside the active region and bright point areas) is: $-$0.00039$\pm$0.00245 Data Numbers s$^{-1}$ and we chose a threshold of $>$ 0.03 Data Numbers s$^{-1}$. Given the known response of XRT to optically thin thermal plasma, we can generate the possible Differential Emission Measure (DEM) curves that can reproduce the observed fluxes in 10 different XRT filters and filter combinations. We chose uncertainties of 3\%, which result from comparing the Monte-Carlo realizations of the full area outlined in Figure 1 with the individual pixel-by-pixel results: the 1 $\sigma$ spread in the two DEM distributions is similar. (This result will be examined in greater detail in a future paper.)

The orange (heavy dashed) curve in Figure 2a shows the {\it optimal median} of the Monte-Carlo iterations generated by the Markov-Chain Monte Carlo (MCMC) based reconstruction algorithm (Kashyap \& Drake 1998). The optimal median is the iteration best matching the median DEM solution in each temperature bin, after discarding non-convergent solutions. The routine fits a locally smoothed DEM curve to the data by comparing the predicted to the observed fluxes and modifying the solution randomly to obtain new realizations. The smoothing scale varies with temperature to account for changes in the information content available as codified in the filter response curves. The new realizations are kept or discarded according to the Metropolis criterion based on changes in the $\chi^2$ values at each step, resulting in a Markov-Chain that explores the parameter space efficiently. This results in a sample of solutions that is representative of the actual probability distribution of the DEM (see e.g., Smith \& Roberts 1993; Casella 1996). Many of the details and advantages of this method, including DEM dynamic range, uncertainties, and smoothing, were discussed by Schmelz, Kashyap \& Weber (2007) and references therein. The predicted-to-observed flux ratio for each filter is listed in Table 1.

The yellow (heavy solid) curve in Figure 2a  is generated with the same XRT data, but using xrt\_dem\_iterative2.pro. Testing and validation of the method with synthetic data leading up to the launch of Hinode were presented by Golub et al. (2004) and Weber et al. (2004). The routine employs a forward-fitting approach where a DEM is guessed and folded through each response to generate predicted fluxes. This process is iterated to reduce the $\chi^2$ between the predicted and observed fluxes. The DEM function is interpolated using several spline points, which are directly manipulated by mpfit.pro, a well-known and much-tested IDL routine that performs a Levenberg-Marquardt least-squares minimization (Levenberg 1944; Marquardt 1963). There are N$_i$ - 1 splines, representing the degrees of freedom for N$_i$ observations. This routine uses Monte-Carlo iterations to estimate errors on the DEM solution (grey dashes in Figure 2a). For each iteration, the observed flux in each filter was varied randomly and the program was run again with the new values. The distribution of these variations was Gaussian with a centroid equal to the observed flux and a width equal to the uncertainty (3\%). The predicted-to-observed flux ratios are also listed in Table 1.  We are pleased to find that the two methods are in excellent agreement: the xrt\_dem\_iterative2.pro results are within 1$\sigma$ of the MCMC results for five of the XRT filters and within 2$\sigma$ for the remaining filters. 

To evaluate the quality of DEM reconstruction that can be achieved with XRT, we have compared the results of xrt\_dem\_iterative2.pro with a known input model. In each case, a nominal observation is calculated for a given DEM model, and then the procedure samples the observation 100 times by including random photon noise at the 3\% level. The distribution of calculated DEM curves (relative to the known DEM model) indicates the accuracy and robustness of the analysis method. Figure 3 (top) shows the results for the isothermal cases: Log T $=$ 6.1 and 6.4.  Each plot shows the input delta function (solid line), the 100 Monte-Carlo realizations (grayscale), and the median of all realizations for each temperature bin (diamonds). The results use all nine filters and demonstrate that the routine can determine the correct result by placing most of the power in the temperature bin corresponding to the delta function model, with a smaller amount placed in the adjacent bins. Recall that the fitted DEM is a spline curve through several knots; although this is not the best function for fitting a delta function, our results are still quite good. For a multi-thermal example, we analyzed the active region model from CHIANTI (v.5.2.1; Dere et al. 1997; Landi et al. 2006). The results suggest how well the input DEM can be reconstructed as a function of the number of observing channels used. Figure 3 (bottom) shows the model AR DEM (solid line with two humps). The first panel uses only the four thickset XRT filters. It can determine the presence of the hotter peak, as indicated by the convergence of the median fit to the model DEM curve, but fails to detect the cooler material. The last panel uses eight filters. The cooler model component is detected, but not accurately represented, which is consistent with the results from Figure 2a where the Monte-Carlo iterations begin to diverge at low temperatures.

The DEM realizations seen in Figure 2a are constrained by 10 XRT filters and filter combinations available for the active region core. The resulting curves show the expected component at Log T $=$ 6.3-6.5, but there is also a high-temperature component associated with this emission that is down by almost three orders of magnitude from the main peak. This component is exactly the type of feature that we would expect if the active region were heated by nanoflares. The lower emission measure explains why it has not been detected by other instruments -- the {\it dynamic range} of a typical DEM curve is approximately two orders of magnitude. Any features beyond this range would, in general, not be trusted. With XRT and its high sensitivity to hot plasma, however, this restriction does not apply. The peak filter responses span almost four orders of magnitude, so the DEM dynamic range is substantially greater than previous instruments. One important point, however, is that the observations do not require that the DEM curve be
bimodal. There is a possibility (see grey Monte-Carlo iterations) that the DEM decreases monotonically beyond Log T $=$ 6.4. This has implications for the possible distribution of nanoflare energies (see below). We also note that all DEM results depend heavily on instrument calibration. A lot of effort has gone into determining the XRT filter thicknesses (Narukage et al. 2009), but problems may still remain. One of the advantages of DEM over traditional filter ratio methods is that these results are not as sensitive to an uncertainty in one particular filter. For example, both DEM methods used here indicate that the predicted flux for the Thick\_Al filter is too high, a problem that could be related to the adopted thickness of the filter. Small changes to the filter thickness change the amount of predicted flux (and therefore the goodness of fit) but not the DEM shape.

One obvious concern is that the high temperature plasma revealed in these DEM curves might result from the GOES A7 flare from the left side of AR 10955. The thickest filters, however, were observed well before the flare (see Table 1), thus reducing the likelihood of significant contamination. To address any residual concern, we have generated synthetic spectra using the ch\_ss.pro routine available in CHIANTI. We have chosen the wavelength range of 13-14 \AA\ because it contains several high temperature iron lines that can be used as a discriminator between flaring and quiescent plasma. The yellow DEM curve from Figure 2a (the orange curve gets similar results) was used to produce the dotted spectrum seen in Figure 4. The Ne IX triplet lines (13.45, 13.55, 13.70 \AA) with a peak formation temperature of Log T $=$ 6.2 as well as an Fe XVII line (13.82 \AA; Log T $=$ 6.6) are clearly visible. These lines are observed routinely from quiescent active regions (Schmelz et al. 1996), but there is no hint of the hotter Fe XIX line (13.52 \AA; Log T $=$ 6.9). The solid spectrum was generated with a DEM from the decay phase of a GOES B8 flare (Reeves et al.  2009). The strongest Fe XIX line is now obvious, as are several smaller Fe XIX and Fe XX flare lines which are all missing from the active region core spectrum. These results indicate that not only is the hot DEM component observed by XRT consistent with earlier observations from spectrometers, but more importantly, it is unlikely to have resulted from the flare. (Note: direct comparison of the AR core and A7 flare spectra would have been optimal, but lack of the thick filter observations during the flare prevented accurate flare DEM analysis; see Table 1 for the exact times.)

We have modeled the observed DEM curve with nanoflares using the EBTEL hydrodynamic simulation code (Klimchuk, Patsourakos \& Cargill 2008). Our goal at this time is not to accurately reproduce the details of the curve, but rather to demonstrate that impulsive heating can explain the overall shape. We therefore make a number of simplifying assumptions. First, we assume the observed volume outlined in Figure 1 is comprised of unresolved loop strands with a half length of 60 Mm. Second, we assume that there are just two types of nanoflares. Most of the strands are heated by weak nanoflares that repeat every $4.0\times10^3$ s, while 0.3\% of the strands are heated by much stronger nanoflares that repeat every $3.0\times10^5$ s.  The model DEM curve is obtained by time averaging the weak and strong nanoflare simulations over these durations and combining in the appropriate proportion.  The nanoflares have a triangular heating profile with a duration of 500 s and amplitude that corresponds to a time-averaged energy flux through the footpoints of $5.625\times10^6$ erg cm$^{-2}$ s$^{-1}$ and $3.0\times10^7$ erg cm$^{-2}$ s$^{-1}$ for the weak and strong events, respectively. This time average includes the long interval between successive nanoflares. Note that the canonical observed radiative loss rate from active regions is $10^7$ erg cm$^{-2}$ s$^{-1}$ (Withbroe \& Noyes 1977). In addition to the nanoflares, there is a very weak steady background heating that corresponds to a static equilibrium with a temperature of 0.45 MK. All strands cool fully (i.e., to 0.45 MK) before being reheated by the next nanoflare, but the plasma drains slowly enough that the density in the weak nanoflare strands never drops below 15 times the equilibrium density. 

The DEM distribution predicted by the model is given in Figure 2b. The solid curve is from the coronal part of the strands, while the dashed curve includes both the corona and the transition region footpoints (moss). The agreement with the observed curve in Figure 2a is remarkably good given the simplicity of the model. It is clear that nanoflare heating is consistent with the observations. Note that XRT does not provide significant constraints for plasma cooler than about 1 MK (for our filter set), so disagreement with this part of the curve is not meaningful. 

Nonequilibrium ionization will reduce the high temperature DEM derived from observations relative to the actual values.  However, we do not expect the effect to be large for 500 s nanoflares like those in our simulations (Bradshaw \& Cargill 2006; Reale \& Orlando 2008). The temperature cooling timescale in the strong nanoflare simulation is always greater than 500 s.

One can ask whether steady heating is also consistent with the observations. It seems likely that a suitable distribution of steady heating could reproduce the DEM curve. However, the extreme level of steady heating required to maintain the hottest stands is highly implausible. For example, an energy flux of several times $10^9$ erg cm$^{-2}$ s$^{-1}$ is required to produce a static equilibrium at 20 MK. This is more than two orders of magnitude greater than the canonical value. If the energy for the heating is provided by the stressing of the magnetic field by footpoint motions, then a steady photospheric velocity of order 100 km s$^{-1}$ is required. Such a flow is completely unsupported by observations. Wave heating is also implausible. Even with a wave speed as large as 1000 km s$^{-1}$, the fluctuation velocity would need to be of order 1000 km s$^{-1}$ to provide the necessary energy flux (assuming that such waves could dissipate fully in the low corona, which is unlikely).  The corresponding displacement amplitude is totally unrealistic for wave periods $>$ 10 s.  Thus, we conclude that while the cooler plasma in the DEM distribution could be due to either impulsive or steady heating, the hotter plasma can only be a result of nanoflares. 

\section{Conclusions}

The DEM distribution generated for a quiescent active region core shows a high-temperature component with significant hot (Log T $>$ 7.0) emission measure. XRT is sensitive to this plasma because it has a greater DEM dynamic range (almost four orders of magnitude) than any earlier solar X-ray imager or spectrometer. The hot plasma result is consistent with observations from spectrometers in that the synthetic spectrum generated using the XRT active region core DEM shows no Fe XIX lines, which would indicate flaring plasma. The EBTEL code was used to successfully model the result. For these reasons, we suggest that this might be the first convincing evidence of a nanoflare signature.

This work was inspired by a talk given by Fabio Reale at the September 2008 Hinode-2 Meeting in Boulder, CO. We would like to thank Fabio, as well as Paola Testa, John Raymond, and Nancy Brickhouse for helpful discussions. Hinode is a Japanese mission developed and launched by ISAS/JAXA, with NAOJ as domestic partner and NASA and STFC (UK) as international partners. It is operated by these agencies in co-operation with ESA and the NSC (Norway). Solar physics research at the University of Memphis is supported by a Hinode subcontract from NASA/SAO as well as NSF ATM-0402729.

{}

\clearpage

\begin{deluxetable}{lcccccc}
\tabletypesize{\scriptsize}
\tablewidth{0pt}
\tablecaption{ XRT Data from 2007 May 13}
\tablehead{
\colhead{Filter} & \colhead{Time} & \colhead{Long Exposure} & \colhead{Time} & \colhead{Short Exposure} & \colhead{Pred/Obs Flux } & \colhead{Pred/Obs Flux } \\
\colhead{} & \colhead{(UT) } & \colhead{ (sec) } & \colhead{(UT) } & \colhead{(sec) } & \colhead{MCMC} & \colhead{((xrt\_dem\_iterative2))} 
}
\startdata
Al\_mesh 	&18:04:43	&4.10 sec 		&18:03:47	&0.18 sec		&1.38$\pm$.04	&1.33$\pm$.04\\
Al\_poly 	&18:17:06	&5.80 sec 		&18:16:38	&0.51 sec		&------$^1$		&------$^1$\\
C\_poly 	&18:10:00	&8.20 sec 		&18:09:01	&0.51 sec		&1.17$\pm$.03	&1.12$\pm$.03\\
Ti\_poly	&18:03:27	&8.20 sec 		&18:03:17	&0.51 sec		&0.86$\pm$.02	&0.82$\pm$.02\\
Al\_poly-Ti	&18:06:33	&16.4 sec 	&18:05:23	&1.45 sec		&0.86$\pm$.02	&0.81$\pm$.02\\
C-Ti		&18:08:17	&16.4 sec 		&18:07:17	&1.03 sec		&0.80$\pm$.02	&0.75$\pm$.02\\
Be\_thin 	&18:12:31	&23.1 sec 		&18:11:47	&1.03 sec		&1.47$\pm$.04	&1.37$\pm$.04\\
Be\_med	&18:14:06	&46.3 sec 		&18:13:13	&2.05 sec		&1.16$\pm$.03	&1.09$\pm$.03\\
Al\_med	&18:15:39	&46.3 sec 		&18:15:04	&2.05 sec		&1.26$\pm$.03	&1.18$\pm$.03\\
Al\_thick	&18:02:19	&46.3 sec 		&18:01:54	&16.4 sec		&1.99$\pm$.07	&1.87$\pm$.07\\
Be\_thick	&18:00:37	&65.5 sec 		& ------	&   ------			&1.09$\pm$.03	&1.03$\pm$.03\\
\enddata
\tablenotetext{1}{Filter not used due to concerns about contamination by atmospheric absorption (near earth eclipse)}
\end{deluxetable}

\clearpage

\begin{figure}
\figurenum{1}
\plotone{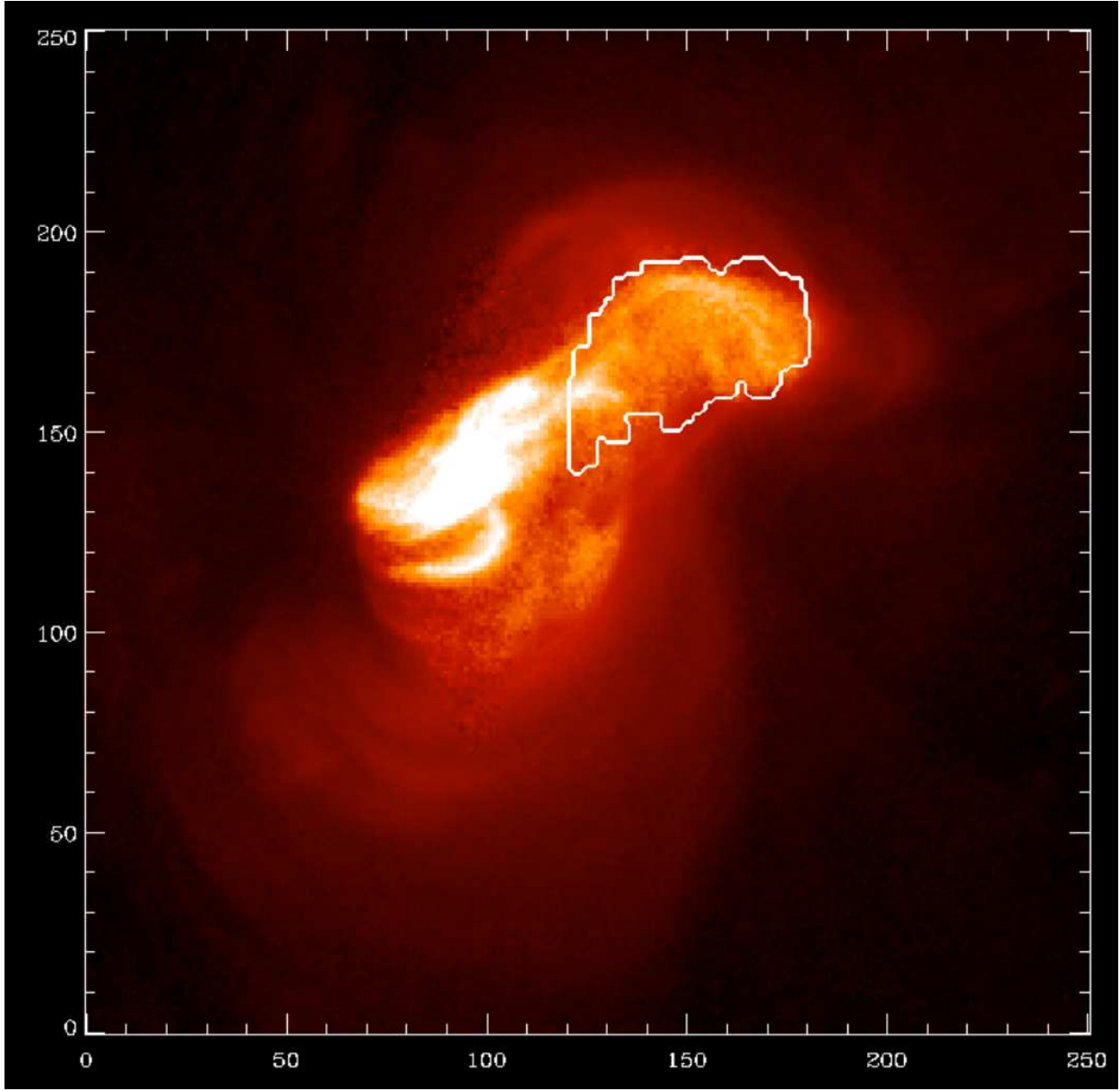}
\caption{Composite long/short exposure XRT image of AR 10955 made with the Ti\_poly filter. The DEM analysis used contiguous pixels within the outlined area which had $>$0.03 Data Numbers s$^{-1}$ in Be\_thick.
}
\end{figure}

\clearpage

\begin{figure}
\figurenum{2a}
\plotone{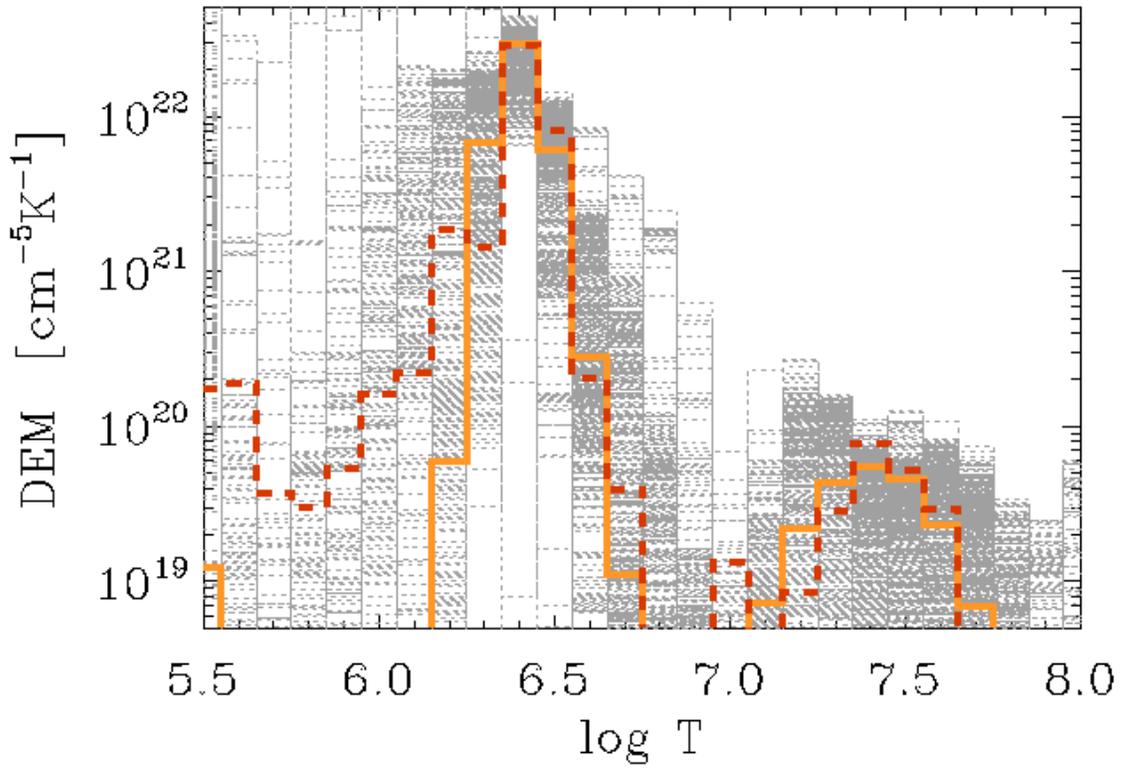}
\caption{Optimal median DEM solutions (see text) from xrt\_dem\_iterative2 (yellow, solid) and MCMC (orange, dashed); the minimum $\chi^2$ solutions are similar. The solution sets from 200 different Monte-Carlo realizations are overplotted as dotted histograms. 
}
\end{figure}

\clearpage

\begin{figure}
\figurenum{2b}
\plotone{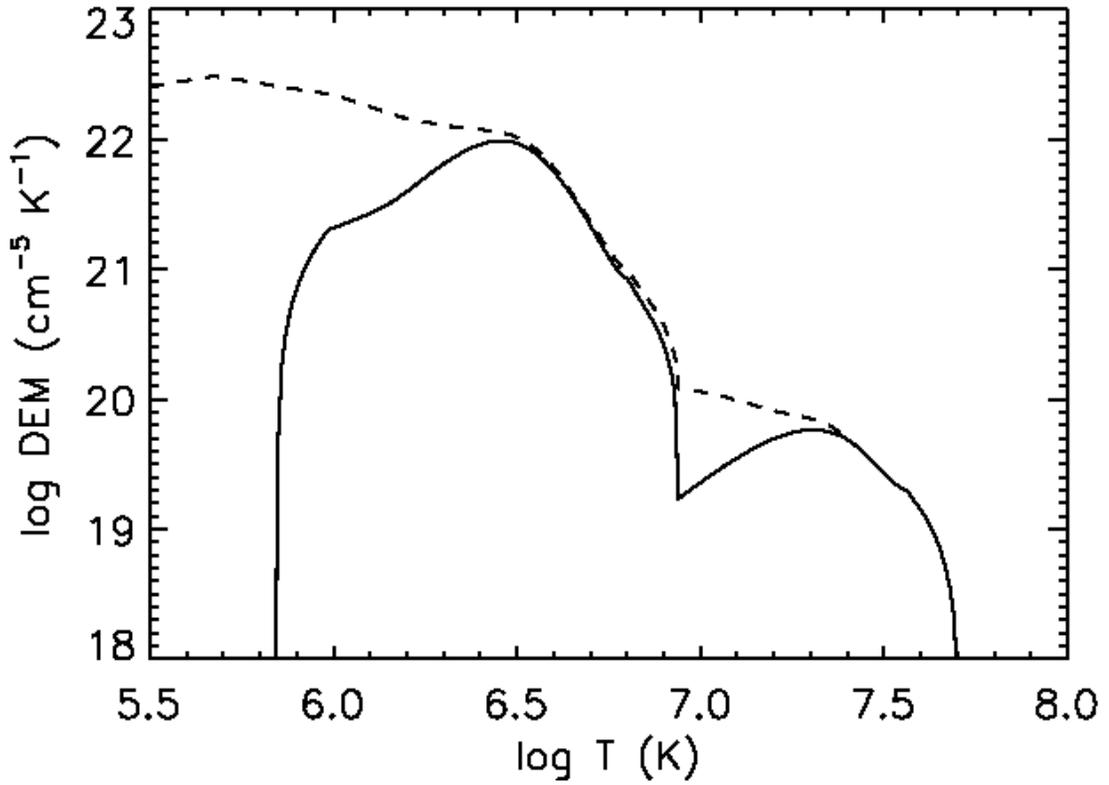}
\caption{Theoretical DEM curves generated from a simple composite of two nanoflare simulations that have been averaged over time and scaled to match the observations. The solid curve includes only the coronal emission where the dashed curve adds the footpoint (moss) emission.
}
\end{figure}

\clearpage

\begin{figure}
\figurenum{3}
\plotone{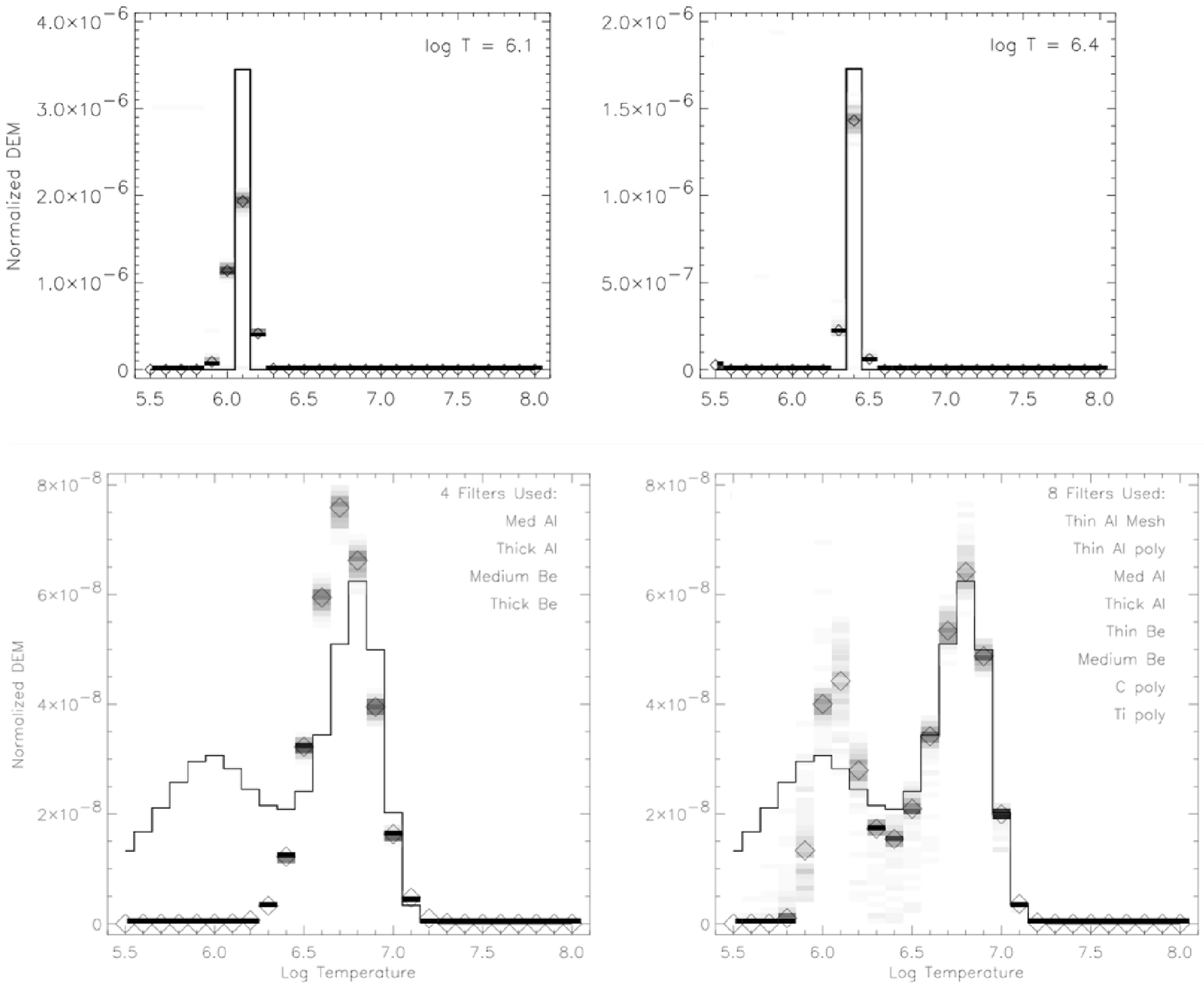}
\caption{Single-T DEMs with Log T $=$ 6.1 and 6.4 (top) and multi-thermal DEM (bottom). The model DEM is shown as a solid line; the grayscale shading indicates the density of 100 Monte Carlo estimates; and the diamonds indicate the median fit.
}
\end{figure}

\clearpage

\begin{figure}
\figurenum{4}
\plotone{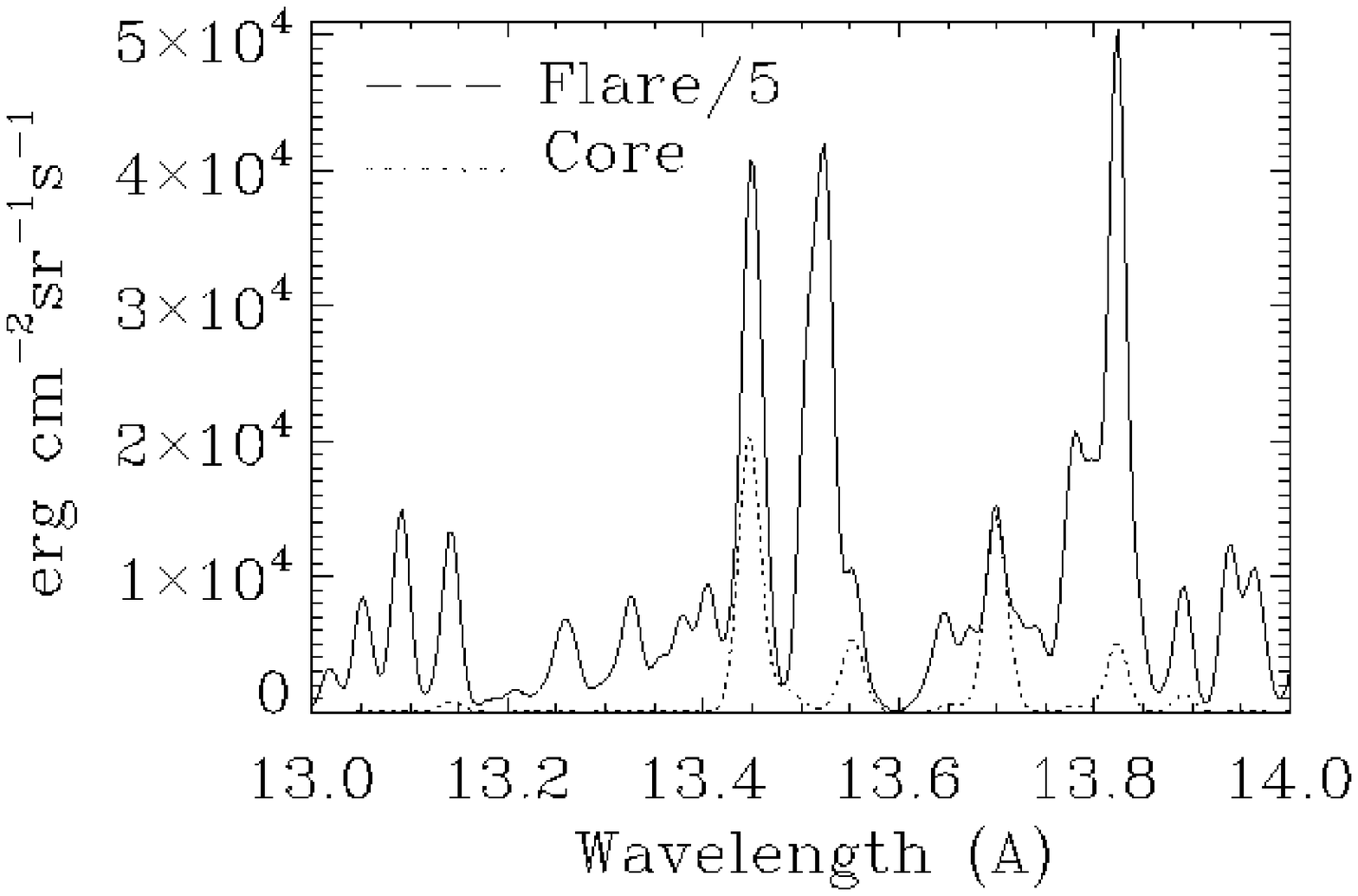}
\caption{Synthetic spectra generated from CHIANTI using the XRT active region core DEM (dot) and the B8 flare DEM (solid) from Reeves et al. ( 2009). The first spectrum is dominated by the Ne IX triplet lines at 13.45, 13.55, and 13.70 \AA\ as well as an Fe XVII line at 13.82 \AA. The flare spectrum is divided by five for comparison. It shows a strong Fe XIX line at 13.52 \AA\ as well as other flare lines.
}
\end{figure}


\begin{thebibliography}{}

\bibitem[]{} Bradshaw, S. J. \& Cargill, P. J.  2006, \aap, 458, 987
\bibitem[]{} Casella, G. 1996, Test, 5, 249
\bibitem[]{} Culhane, J.L et al. 1991, \solphys, 136, 89
\bibitem[]{} Dere, K.P., Landi, E., Mason, H.E., Monsignori Fossi, B.C., Young, P.R. 1997, \aaps, 125, 149
\bibitem[]{} Fludra, A. \& Schmelz, J.T. 1999, \aap, 348, 286
\bibitem[]{} Golub, L. et al. 2007, \solphys, 2007, 243, 63
\bibitem[]{} Golub, L., Deluca, E.E., Sette, A., Weber, M. 2004, ASP Conference Series, Tokyo, Japan. Eds. T. Sakurai and T. Sekii, 325, 217
\bibitem[]{} Kashyap, V. \& Drake, J.J. 1998, \apj, 503, 450
\bibitem[]{} Klimchuk, J.A. 2006, \solphys, 234, 41
\bibitem[]{} Klimchuk, J.A., Patsourakos, S., Cargill, P.J. 2008, \apj, 682, 1351
\bibitem[]{} Kosugi, T. et al. 2007, \solphys, 243, 3
\bibitem[]{} Landi, E., Del Zanna, G., Young, P.R., Dere, K.P., Mason, H.E., Landini, M. 2006, \apjs, 162, 261
\bibitem[]{} Lemen, J.R., Bentley, R.D., Sylwester, J. 1986, Adv. Spa. Res, 6, 245
\bibitem[]{} Levenberg, K. 1944, The Quarterly of Applied Mathematics 2, 164
\bibitem[]{} Marquardt, D. 1963, SIAM Journal on Applied Mathematics 11, 431
\bibitem[]{} Narukage, N. et al.  2009, in prep.
\bibitem[]{} Parker, E.N. 1983, \apj, 264, 642
\bibitem[]{} Parker, E.N. 1988, \apj 330, 474
\bibitem[]{} Reale, F. \& Orlando, S.  2008, \apj, 284, 715
\bibitem[]{} Reale, F. et al. 2007, Science, 318, 1582
\bibitem[]{} Reeves, K. et al.  2009, in prep.
\bibitem[]{} Schmelz, J.T. 1993, \apj, 408, 373
\bibitem[]{} Schmelz, J.T., Kashyap, V. L. \& Weber, M.A. 2007, \apjl, 660, L157
\bibitem[]{} Schmelz, J.T., Saba, J.L.R., Ghosh, D. \& Strong, K.T. 1996, \apj, 473, 519
\bibitem[]{} Smith, A.F.M. \& Roberts, G.O. 1993, Journal of the Royal Statistical Society, Series B, 55, 3
\bibitem[]{} Sterling, A. 1997, Geo Res. Let., 24, 2263
\bibitem[]{} Teriaca, L., Falchi, A., Falciani, R., Cauzzi, G., Maltagliati, L. 2006, \aap, 455, 1123
\bibitem[]{} Tsuneta, S. et al. 1991, \solphys, 136, 37
\bibitem[]{} Wang, T.J. Innes, D.E., Solanki, S.K. 2006, \aap, 455, 1105
\bibitem[]{} Watanabe, T et al. 1995, \solphys, 157, 169
\bibitem[]{} Weber, M.A., DeLuca, E.E., Golub, L. \& Sette, A.L. 2004, Proc. IAU Symp. 223, Multi- 
Wavelength Investigations of Solar Activity, 321
\bibitem[]{} Withbroe, G.L.\& Noyes, R.W. 1977, \araa, 15, 363


\end{thebibliography}
\end{document}